\documentclass[letterpaper, 10 pt, conference]{ieeeconf}  
\IEEEoverridecommandlockouts                              
\usepackage{graphics} 
\usepackage{epsfig} 
\usepackage{amsmath} 
\usepackage{amssymb}  

\usepackage{graphicx}

\usepackage{amsthm}

\newtheorem{definition}{\bf Definition}
\newtheorem{theorem}{\bf Theorem}
\newtheorem{proposition}{\bf Proposition}

\newtheorem{remark}{\bf Remark}

\newcommand{\prob}[1]{\mathrm{Pr}\left[#1\right]}
\newcommand{\expect}[1]{\mathbb{E}\left[#1\right]} 
\newcommand{\tr}[1]{\mathrm{tr}\left(#1\right)}

\newcommand{\betabar}{\beta_{\max}}
\newcommand{\gammabar}{\gamma_{\max}}
\newcommand{\tildebetabar}{\tilde{\beta}_{\max}}

\usepackage{setspace}

\usepackage{tikz}
 \usetikzlibrary{arrows,patterns,mindmap,backgrounds,shadows}
\usetikzlibrary{backgrounds}
\usetikzlibrary{shapes,arrows,chains}

\usepackage{siunitx}

\usepackage{cite}
\usepackage{hyperref}
\usepackage{comment}

\begin{document} 
\begin{minipage}{\textwidth}\ \\[18pt] \\ \\
         \copyright 2023 IEEE.  Personal use of this material is  permitted.  Permission from IEEE must be obtained for all other uses, in  any current or future media, including reprinting/republishing this material for advertising or promotional purposes, creating new  collective works, for resale or redistribution to servers or lists, or  reuse of any copyrighted component of this work in other works.
     \end{minipage}

\title{{On stochastic MPC formulations with closed-loop guarantees:\\
Analysis and a unifying framework - extended version}}
\author{Johannes K\"ohler$^\star$, Ferdinand Geuss$^\star$, Melanie N. Zeilinger%
\thanks{$^\star$Johannes K\"ohler and Ferdinand Geuss contributed equally to this paper.}
\thanks{Institute for Dynamic Systems and Control, ETH Zürich, Zürich CH-8092, Switzerland}
\thanks{Johannes K\"ohler was supported by an ETH Career Seed Award funded through the ETH Z\"urich Foundation and Swiss National Science Foundation under NCCR Automation (grant agreement 51NF40 180545).}}

\maketitle
\begin{abstract} 
We investigate model predictive control (MPC) formulations for linear systems subject to i.i.d. stochastic disturbances with bounded support and chance constraints.  
Existing stochastic MPC formulations with closed-loop guarantees can be broadly classified in two separate frameworks: 
i)~using robust techniques; ii)~feasibility preserving algorithms. 
We investigate two particular MPC formulations representative for these two frameworks called \textit{robust-stochastic} MPC and \textit{indirect feedback} stochastic MPC.
We provide a qualitative analysis, highlighting intrinsic limitations of both approaches in different edge cases. 
Then, we derive a unifying stochastic MPC framework that naturally includes these two formulations as limit cases.  
This qualitative analysis is complemented with numerical results, showcasing the advantages and limitations of each method. 

\end{abstract}  
\setstretch{0.99}

\section{Introduction}
Model predictive control (MPC) is an optimization-based control strategy that can ensure satisfaction of state and input constraints~\cite{rawlings2017model}. 
In order to account for disturbances, robust or stochastic MPC formulations can be utilized~\cite{Kouvaritakis2016textbook}.

\subsubsection*{Stochastic MPC (SMPC)}
SMPC formulations consider chance constraints, i.e.,
constraint satisfaction with some user-chosen probability, to avoid the conservatism of worst-case robust formulations~\cite{mesbah2016stochastic,Farina2016soverview}. 
The reformulation of (standard) linear stochastic optimal control problems as a deterministic quadratic program (QP) has been largely solved in the literature, e.g., using a constraint tightening based on analytical reformulations or scenario-based approximation~\cite{mesbah2016stochastic,Farina2016soverview}. 
However, the development of receding horizon SMPC algorithms turns out to be non-trivial. 
 In particular, the stochastic formulation explicitly allows for a certain probability of constraint violation. Hence, the optimization problem can become infeasible during closed-loop operation. 
In the SMPC literature, two main frameworks have emerged 
that address this issue and yield recursive feasibility, performance, and chance constraint satisfaction: 
utilizing \textit{robust} techniques or redesigning the MPC algorithm.

\subsubsection*{Robust techniques} 
A natural solution to this problem is \textit{robustly} ensuring
recursive feasibility~\cite{cannon2010stochastic,KouvaritakisCannon2010rs-mpc,cannon2012stochastic,korda2011strongly,Lorenzen2017tightening,kerz2023dataSMPC,Kouvaritakis2016textbook}. 
In particular, one can use a constraint tightening composed of primarily robust bounds where only the first step is done stochastically, which ensures that the shifted candidate solution is feasible for any possible disturbance~\cite{cannon2010stochastic,KouvaritakisCannon2010rs-mpc,cannon2012stochastic}. 
Due to the primary reliance on robust constraint tightening, we refer to this approach as \textit{robust-stochastic} MPC. 
A less conservative approach is used in~\cite{korda2011strongly,Lorenzen2017tightening,kerz2023dataSMPC} by directly enforcing robust recursive feasibility using the robust control invariant set. 
 
\subsubsection*{Feasibility preserving algorithms} 
In order to address possibly unbounded (e.g., Gaussian) disturbances, a second SMPC framework has emerged over the last decade which modifies standard MPC algorithms to ensure recursive feasibility~\cite{farina2013probabilistic,farina2015approach,farina2016model,li2021distributionally,pan2022towards,Hewing2018recovery_mechanism,mark2021stochastic,li2022corrigendum,Koehler2022interpolating,schluter2022stochastic,gruner2022recursively,Hewing2020indirect,mayne2016robust,Hewing2020direct_vs_indirect,mark2021data,hewing2019scenario,muntwiler2022lqg,wang2021recursive}.
Early work in~\cite{farina2013probabilistic} suggested to only use the new measured state if this retains recursive feasibility, which was also adopted/extended in~\cite{farina2015approach,farina2016model,li2021distributionally,pan2022towards}.
More recently, in~\cite{Hewing2018recovery_mechanism}, it was shown that a modified version of this approach does in fact provide the desired closed-loop guarantees, which was subsequently utilized and extended in~\cite{mark2021stochastic,li2022corrigendum,Koehler2022interpolating,schluter2022stochastic,gruner2022recursively}. 
\textit{Indirect feedback SMPC}~\cite{Hewing2020indirect} provides a simpler way to address feasibility by making the nominal state (and hence the feasible set) completely independent of the realized disturbances.
Similar ideas were advocated in~\cite{mayne2016robust}, and extensions of this idea can be found in~\cite{Hewing2020direct_vs_indirect,mark2021data,hewing2019scenario,muntwiler2022lqg}, compare also \cite{wang2021recursive} for a comparable approach. 
\subsubsection*{Contribution}
In this paper, we contrast these two available SMPC frameworks by carving out their relationships and individual trade-offs. 

Since these frameworks
were introduced for different problem setups (bounded vs. unbounded support), there exists no\footnote{%
In fact, \cite{Hewing2020direct_vs_indirect,muntwiler2022lqg,Koehler2022interpolating} provide some recent comparisons between SMPC schemes. However, due to consideration of unbounded Gaussian noise, these discussions and comparisons do not consider the robust-stochastic MPC schemes~\cite{cannon2010stochastic,KouvaritakisCannon2010rs-mpc,cannon2012stochastic,korda2011strongly,Lorenzen2017tightening,Kouvaritakis2016textbook}.} analysis regarding their benefits and limitations. 
We consider two specific SMPC schemes that are representative of these two frameworks: the robust-stochastic MPC (RS-MPC) from~\cite[Chap.~8.1]{Kouvaritakis2016textbook} and the indirect feedback SMPC (IF-SMPC) from~\cite{Hewing2020indirect}. 
We provide a qualitative analysis by showing edge cases where either one of the two approaches is guaranteed superior. 
This analysis is complemented with numerical simulations, which show that the quantitative difference between these formulations can be significant.

As a separate contribution, we derive a novel SMPC framework, called multi-step SMPC (MS-SMPC), which naturally unifies these two SMPC formulations.
In particular, the difference between RS-MPC and IF-SMPC is on which state the chance constraints are conditioned: the current measured state or the initial state.
The proposed SMPC formulation conditions the chance constraints on a state (up to) $M$ time-steps in the past with some user-chosen $M$. 
For $M\in\{1,\infty\}$, this naturally recovers the two existing SMPC formulations. 

\subsubsection*{Outline} 
We first present the problem setup in Section~\ref{sec:setup} and introduce preliminaries regarding SMPC (Sec.~\ref{sec:tightening}). 
Section~\ref{sec:SMPC} presents the two SMPC formulations, RS-MPC \& IF-SMPC, and provides a qualitative analysis regarding benefits and limitations. 
 Then, we derive a unifying framework (Sec.~\ref{sec:multiStep}), provide a numerical comparison (Sec.~\ref{sec:numerical_comparison}),
and end the paper with some conclusions (Sec.~\ref{sec:conclusion}). 
The appendix contains the theoretical proof for the unifying SMPC framework (App.~\ref{app:proof}) and additional details regarding the computation of the constraint tightening (App.~\ref{app:tightening_infinite_horizon}).

\subsubsection*{Notation}
The set of integers in an interval $[a,b]$ is denoted by $\mathbb{I}_{[a,b]}$.
The modulo operator for $k,M\in\mathbb{I}_{\geq 0}$ is denoted by $\mathrm{mod}(k,M)\in\mathbb{I}_{[0,M-1]}$. 
 For vectors $a,b\in\mathbb{R}^n$, $a\leq b$ denotes an element-wise inequality. 
 We denote a diagonal matrix with diagonal elements $a_i\in\mathbb{R}$, $i\in\mathbb{I}_{[1,n]}$ by $\text{diag}(a_1,\dots a_n)\in\mathbb{R}^{n\times n}$. 
A vector of ones is denoted by $\mathbf{1}$. 
The trace of a square matrix $A$ is denoted by $\tr{A}$.
The $j$-th row of a matrix $F\in\mathbb{R}^{r\times n}$ is denoted by $F_{(j)}\in\mathbb{R}^{1\times n}$. 
For a vector $x\in\mathbb{R}^n$ and a positive semi-definite matrix $Q\in\mathbb{R}^{n\times n}$, we abbreviate $\|x\|_Q^2:=x^\top Q x$. 
The expected value of a stochastic variable $w$ is denoted by $\expect{w}$. 
The probability of an event $A$ is denoted by $\prob{A}$. 
By $c_{i|k}$, $k\in\mathbb{I}_{\geq 0}$, $i\in\mathbb{I}_{\geq 0}$ we denote a prediction of a variable $c$ at time $k$, $i$ steps in the future. 
\section{Problem setup}
\label{sec:setup}
We consider linear systems of the form
\begin{equation}
    x(k+1) = A x(k) + B u(k) + D w(k) 
    \label{eq:linear_system}
\end{equation}
with state $x(k)\in\mathbb{R}^n$, input $u(k)\in\mathbb{R}^m$, and additive disturbance $w(k)\in\mathbb{R}^q$. 
The disturbances are  independent and identically distributed (i.i.d.)  with zero-mean and variance $\Sigma_{\mathrm{w}}\in\mathbb{R}^{q\times q}$. 
In addition, we assume that the disturbances are bounded, i.e., $w(k)\in\mathcal{W}$, $\forall k\in\mathbb{I}_{\geq 0}$ with some known compact set $\mathcal{W}$.  
The system is constrained by individual half-space chance constraints 
\begin{equation}
\label{eq:prob_constraint}
    \prob{ H_{\mathrm{x},(j)} x(k) \leq 1 } \geq p_j , ~j \in \mathbb{I}_{[1,r_{\mathrm{x}}]}
\end{equation}
with some probability level $p_j\in(0,1]$.
Furthermore, we have 
hard polytopic input constraints of the form
\begin{equation}
\label{eq:constraint_input}
    H_{\mathrm{u},(j)} u(k) \leq 1 , ~j \in \mathbb{I}_{[1,r_{\mathrm{u}}]}. 
\end{equation}
The constraints~\eqref{eq:prob_constraint}--\eqref{eq:constraint_input} can be equivalently formulated into mixed state and input constraints 
\begin{equation}
    \prob{ F_{(j)} x(k) + G_{(j)} u(k) \leq 1 } \geq p_{j} , ~j \in \mathbb{I}_{[1,r]} 
    \label{eq:general_stochastic_constraint}
\end{equation}
with $r = r_{\mathrm{x}} + r_{\mathrm{u}}$ 
and $p_j = 1$ 
for $j\in\mathbb{I}_{[r_{\mathrm{x}}+1,r_{\mathrm{x}}+r_{\mathrm{u}}]}$.  Furthermore, we have a convex linear-quadratic stage cost
$\ell(x,u):=\|x\|_Q^2+x^\top q +\|u\|_R^2+u^\top r$
that should be minimized.
The control problem can be summarized by the following stochastic optimal control problem
\begin{subequations}
\label{eq:stoch_opt_control}
\begin{align}
\label{eq:stoch_opt_control_cost}
\min_{u(\cdot)}&~\expect{\sum_{k=0}^\infty \ell(x(k),u(k))}\\
\text{s.t. }& 
\eqref{eq:linear_system},~\eqref{eq:general_stochastic_constraint},~\forall k\in\mathbb{I}_{\geq 0}.
\end{align}
\end{subequations}
\begin{remark}
\label{rk:setup}(Problem setup) 
A large part of the stochastic MPC literature deals with the challenges related to unbounded (e.g., Gaussian) disturbances~\cite{farina2013probabilistic,Hewing2018recovery_mechanism,Hewing2020indirect,Koehler2022interpolating}. 
These approaches typically relax the hard input constraints~\eqref{eq:constraint_input} to chance constraints (although non-trivial methods to consider both exist, cf. \cite{hewing2019scenario}, \cite[Table~2]{mesbah2016stochastic}). 
We consider bounded disturbances to allow for a comparison to the RS-MPC methods in~\cite{Kouvaritakis2016textbook,cannon2010stochastic,KouvaritakisCannon2010rs-mpc,Lorenzen2017tightening,korda2011strongly,kerz2023dataSMPC,cannon2012stochastic}. 
Furthermore, we consider hard input constraints due to their prevalence in practical application and the inability of the approaches in~\cite{Kouvaritakis2016textbook,korda2011strongly,cannon2010stochastic,KouvaritakisCannon2010rs-mpc,Lorenzen2017tightening,kerz2023dataSMPC} to consider chance input constraints (see Rk.~\ref{rk:input_constraints} below regarding the treatment of input constraints).
\end{remark}

\section{Preliminaries - stochastic optimal control}
\label{sec:tightening}
In the following, we discuss how to reformulate Problem~\eqref{eq:stoch_opt_control} into a QP by using a simple input parametrization and offline constraint tightening, as standard in the SMPC literature~\cite{Kouvaritakis2016textbook,Hewing2020indirect}. 
We consider a linear parametrization for the control input\footnote{%
More general disturbance affine feedbacks would require a more complex chance constraint reformulation in Prop.~\ref{prop:exact_stochastic_constraint}, see, e.g.,~\cite{oldewurtel2008tractable}.}
\begin{equation}
    u(k) = K x(k) + c(k) 
    \label{eq:input_parametrisation}
\end{equation}
with a free input $c(k)\in\mathbb{R}^m$ and an offline chosen stabilizing state feedback $K\in\mathbb{R}^{m\times n}$.
This results in dynamics $x(k+1)=\Phi x(k)+Bc(k)+Dw(k)$
with $\Phi:= A + BK$ Schur stable.  
We define nominal dynamics 
\begin{align}
    s(k+1) &= \Phi s(k) + B c(k),~s(0)=x(0) \label{eq:nominal_state_dynamics}
\end{align}
and an error $e(k) = x(k) - s(k)$, which satisfies
\begin{align}
    e(k+1) &= \Phi e(k) + D w(k),~e(0)=0.
    \label{eq:error_dynamics}
\end{align} 
The following proposition formulates the chance constraints as tightened constraints on the nominal state $s$. 
\begin{proposition}(adapted from~\cite[Lemma~8.1, Sec.~3.2]{Kouvaritakis2016textbook})\\
\label{prop:exact_stochastic_constraint} 
    The stochastic constraint \eqref{eq:general_stochastic_constraint} with \eqref{eq:input_parametrisation}, \eqref{eq:nominal_state_dynamics}, \eqref{eq:error_dynamics} for $k\in\mathbb{I}_{\geq 0}$  
    is equivalent to 
    \begin{equation}
        \tilde{F} s(k) + G c(k) \leq \mathbf{1} - \gamma_k,
        \label{eq:exact_stochastic_constraint}
    \end{equation}
    where $\tilde{F}:=F+GK$ and 
    \begin{subequations}
    \label{eq:exact_stochastic_tightening_term}
        \begin{align}
            \gamma_{k,(j)} &= { \min_{\gamma}\gamma } \label{eq:exact_stochastic_tightening_term_1}
            \\ 
            \mathrm{s.t.} &\;\;\; \prob{ \tilde{F}_{(j)} e(k) \leq \gamma} \geq p_{j} ,~
            j \in \mathbb{I}_{[1,r]}.
            \label{eq:exact_stochastic_tightening_term_2}
        \end{align}
    \end{subequations}
    If $p_j = 1$, \eqref{eq:exact_stochastic_constraint}--\eqref{eq:exact_stochastic_tightening_term} simplifies to 
    \begin{align}
        \label{eq:robust_constraint}
        &\tilde{F} s(k) + G c(k) \leq \mathbf{1} - \sum_{l = 0}^{k-1} a_{l}, \\ 
        &a_{l,(j)} = \max_{w \in \mathcal{W}} \tilde{F}_{(j)} \Phi^l D w,~l\in\mathbb{I}_{\geq 0},~j\in\mathbb{I}_{[1,r]}.
        \label{eq:robust_tightening_term}
    \end{align}
\end{proposition}  
Tightening constants $\gamma_k$~\eqref{eq:exact_stochastic_tightening_term} can be  computed/approximated offline~\cite[Sec.~V.A]{Lorenzen2017tightening}, e.g., using the scenario approach, while the constants $a_{k}$ can be computed using linear programming for polytopic $\mathcal{W}$. We note that the computation of constants $\gamma_k$ can be equally posed as the computation of probabilistic reachable sets for the error, compare~\cite{Hewing2018recovery_mechanism,Hewing2020indirect}.
For large horizons $k\geq \overline{k}$, with some $\bar{k}\gg 1$, 
we can set $\gamma_k=\gammabar$, with some constant $\gammabar$ satisfying~\eqref{eq:exact_stochastic_tightening_term_2} for all $k\in\mathbb{I}_{\geq 0}$. 
Details regarding the computation of $\gamma_k$ and $\gammabar$ can be found in Appendix~\ref{app:tightening_infinite_horizon}.

We consider a finite horizon $N\in\mathbb{I}_{\geq 1}$ and optimize over an input sequence $c_{i|k}\in\mathbb{R}^m$, $i\in\mathbb{I}_{[0,N-1]}$, where we set $c_{i|k}=0$ for $i\in\mathbb{I}_{\geq N}$. 
In the considered setting and parametrization, minimizing the expected cost in~\eqref{eq:stoch_opt_control_cost} yields the same minimizer as minimizing the cost of the mean prediction~\cite{Hewing2020indirect}. 
Hence, considering an initial state $x(k)$, 
and a predicted input sequence $c_{i|k}\in\mathbb{R}^m$, $i\in\mathbb{I}_{[0,N-1]}$ over a horizon $N$, we arrive at the following cost:
\begin{align}
\label{eq:cost}
\mathcal{J}(x(k),c_{\cdot|k}):=\sum_{k=0}^{N-1}\ell(\bar{x}_{i|k},\bar{u}_{i|k})+V_{\mathrm{f}}(\bar{x}_{N|k}),
\end{align}
with the mean prediction $\bar{x}_{0|k}=x(k)$, $\bar{u}_{i|k}=c_{i|k}+K\bar{x}_{i|k}$, $\bar{x}_{i+1|k}=\Phi \bar{x}_{i|k}+Bc_{i|k}$, $i\in\mathbb{I}_{[0,N-1]}$~\cite[Thm.~6.1]{Kouvaritakis2016textbook}. 
Furthermore, $V_{\mathrm{f}}(x):=\|x\|_{P_{\mathrm{f}}}^2+x^\top p_{\mathrm{f}}$ is the linear-quadratic terminal cost that accounts for the infinite-horizon tail, i.e.,
\begin{subequations}
\label{eq:terminal_cost}    
\begin{align}
\label{eq:terminal_cost_quad}    
\Phi^\top P_{\mathrm{f}} \Phi+Q+K^\top RK=P_{\mathrm{f}}, \\
\label{eq:terminal_cost_linear}    
\Phi^\top p_{\mathrm{f}}=p_{\mathrm{f}}+q+K^\top r.
\end{align}
\end{subequations}
Finally, the stochastic optimal control problem can be formulated using the following (finite-dimensional) QP 
\begin{subequations}
 \label{eq:stoch_opt_control_deterministic}
\begin{align}
\min_{c_{\cdot|k}}&~ \mathcal{J}(x(k),c_{\cdot |k})\\
\text{s.t. }& s_{0|k}=x(k)\\
&s_{i+1|k}=\Phi s_{i|k}+Bc_{i|k},~i\in\mathbb{I}_{[0,N-1]}\\
 \label{eq:stoch_opt_control_deterministic_tightenedConstraints}
&\tilde{F}s_{i|k}+Gc_{i|k}\leq \mathbf{1}-\gamma_i,~i\in\mathbb{I}_{[0,N-1]}\\
\label{eq:stoch_opt_control_deterministic_terminal}
&s_{N|k}\in\mathcal{X}_{\mathrm{f}}. 
\end{align}    
\end{subequations}
The terminal set constraint~\eqref{eq:stoch_opt_control_deterministic_terminal} enforces~\eqref{eq:stoch_opt_control_deterministic_tightenedConstraints} for $i\in\mathbb{I}_{\geq N}$ with $c_{i|k}=0$, $i\in\mathbb{I}_{\geq N}$ 
and $\mathcal{X}_{\mathrm{f}}=\{s|\tilde{F}\Phi^is\leq \mathbf{1}-\gamma_{N+i},~i\in\mathbb{I}_{\geq 0}\}$,    
 which admits a finite polytopic representation~\cite[Thm.~8.2]{Kouvaritakis2016textbook}. 
Solving Problem~\eqref{eq:stoch_opt_control_deterministic} at time $k=0$ yields a suboptimal but feasible solution to Problem~\eqref{eq:stoch_opt_control}. 
However, an SMPC scheme based on Problem~\eqref{eq:stoch_opt_control_deterministic} would not be recursively feasible (cf.~\cite{mesbah2016stochastic,Farina2016soverview}).   
\begin{definition}(Desired closed-loop properties)
\label{def:SMPC_good}
The closed-loop system resulting from an SMPC scheme should satisfy the following properties:  
\begin{enumerate}
    \item Given initial feasibility, the SMPC formulation is recursively feasible and hence the closed loop is well-defined.
    \item The chance constraints and input constraints~\eqref{eq:prob_constraint}--\eqref{eq:constraint_input}
 are satisfied for all times $k\in\mathbb{I}_{\geq 0}$.
    \item The asymptotic average performance is no worse than applying the linear feedback $u=Kx$, i.e.,
\begin{equation}
\ell_{\mathrm{avg}}:=
\limsup_{T \rightarrow \infty} \expect{\frac{1}{T} \sum_{k = 0}^{T - 1} \ell(x(k),u(k))}\leq \tr{P_{\mathrm{f}}\Sigma_{\mathrm{w}}}.
    \label{eq:performance}
\end{equation}
\end{enumerate}
\end{definition}
In the following, we assume that $\ell$ admits a uniform lower bound\footnote{%
 This condition is invoked to obtain the performance bound~\eqref{eq:performance} from an expected cost decrease in $\mathcal{J}$, see~\cite[Prop.~5]{Koehler2022interpolating}, \cite[Cor.~2]{Hewing2018recovery_mechanism}.}, which holds trivially if $\ell$ is positive definite. 
Next, we discuss two SMPC formulations that meet these requirements.

\section{Analysis of SMPC frameworks}
\label{sec:SMPC}
In this section, we present the robust-stochastic MPC (RS-MPC)~\cite{Kouvaritakis2016textbook} (Sec.~\ref{sec:RS}) and the indirect feedback SMPC~\cite{Hewing2020indirect} (IF-SMPC) (Sec.~\ref{sec:IF}), which are representative for the two SMPC frameworks commonly considered in the literature.
 Then, we provide a qualitative analysis, 
revealing particular shortcomings of either scheme (Sec.~\ref{sec:discussion_qual}).

\subsection{Robust-stochastic MPC}
\label{sec:RS}
First, we present RS-MPC~\cite{Kouvaritakis2016textbook}, which modifies the SMPC formulation~\eqref{eq:stoch_opt_control_deterministic} by using more conservative constraint tightening constants $\beta_i\geq \gamma_i$ that ensure robust recursive feasibility. 
In particular, this SMPC scheme is based on the following optimization problem:
\begin{subequations}
    \begin{align} 
        \min_{c_{\cdot|k}} \;\;\; &\mathcal{J}(x(k),c_{\cdot|k}) \\
        \mathrm{s. t.} \;\;\; & s_{0|k} = x(k) \\
        & s_{i+1|k} = \Phi s_{i|k} + B c_{i|k} ,~ i \in \mathbb{I}_{[0,N-1]} \\
        & \tilde{F} s_{i|k} + G c_{i|k} \leq \mathbf{1} - \beta_i ,~ i \in \mathbb{I}_{[1,N-1]} \label{eq:Robust_Stochastic_MPC_problem_constraints} \\
        & s_{N|k} \in \mathcal{X}_{\mathrm{f}}^{\mathrm{RS}} \label{eq:Robust_Stochastic_MPC_problem_terminal_set_constraint} \\
        & H_{\mathrm{u}} \left( K x(k) + c_{0|k} \right) \leq \mathbf{1}. \label{eq:Robust_Stochastic_MPC_problem_u0_constraint}
    \end{align}
    \label{eq:Robust_Stochastic_MPC_problem}
\end{subequations}
We denote a minimizer by $^\star$.  
In closed loop, we solve Problem~\eqref{eq:Robust_Stochastic_MPC_problem} at each time $k\in\mathbb{I}_{\geq 0}$ and apply $u(k)=Kx+c_{0|k}^\star$  to the system.  
The constraint tightening $\beta_i$ is constructed by treating only the first disturbance stochastically, while the rest is handled robustly: 
\begin{align}
    \beta_i = \gamma_1 + \sum_{l=1}^{i-1} a_l , \quad i\in\mathbb{I}_{\geq 1} .
    \label{eq:robust_stochastic_tightening_term}
\end{align}
Furthermore, stochastic constraints on the state only need to be enforced for $i \geq 1$, since $s_{0|k}=x(k)$ is deterministic. 
Hence, the hard input constraint for $i=0$ is implemented separately~\eqref{eq:Robust_Stochastic_MPC_problem_u0_constraint}.  
A terminal set constraint \eqref{eq:Robust_Stochastic_MPC_problem_terminal_set_constraint} enforces the constraints \eqref{eq:Robust_Stochastic_MPC_problem_constraints} for $i \geq N$ with $c_{i|k} = 0$, i.e., 
\begin{equation}
\mathcal{X}_{\mathrm{f}}^{\mathrm{RS}} = \{ s | \tilde{F} \Phi^i s \leq \mathbf{1} - \beta_{i+N},~i \in \mathbb{I}_{\geq 0}\}.
\end{equation}
Compared to Problem~\eqref{eq:stoch_opt_control_deterministic}, the main difference is the fact that the constants $\gamma_i$ are replaced by more conservative tightenings $\beta_i\geq \gamma_i$, which yields all the desired closed-loop guarantees.  
\begin{theorem}
\label{thm:rs_closedloop}
(adapted from~\cite[Thm.~8.1, Cor.~8.1]{Kouvaritakis2016textbook})
The SMPC scheme based on Problem~\eqref{eq:Robust_Stochastic_MPC_problem} ensures the desired closed-loop properties in Definition~\ref{def:SMPC_good}. 
\end{theorem}
The key insight in this approach is that the tightening $\beta_i$ turns out to be the least restrictive tightening to \textit{robustly} ensure recursive feasibility with the standard (shifted) candidate solution~\cite[Theorem~8.1]{Kouvaritakis2016textbook}. In general, the approach tries to robustly ensure satisfaction of the chance constraints, hence the name \textit{robust-stochastic MPC}.

\subsection{Indirect feedback SMPC}
\label{sec:IF}
An alternative SMPC method with closed-loop guarantees is IF-SMPC \cite{Hewing2020indirect}. Contrary to the SMPC formulation \eqref{eq:stoch_opt_control_deterministic}, the nominal state $s$ is not reset to the measured state $x(k)$ at each time $k$, but follows the nominal dynamics~\eqref{eq:nominal_state_dynamics}. 
The corresponding optimization problem is provided below:
\begin{subequations}
    \begin{align}
        \min_{c_{\cdot|k}} \;\;\; & \mathcal{J}(x(k),c_{\cdot|k}) \\
        \label{eq:indirect_feedback_SMPC_problem_initialState}
        \mathrm{s. t.} \;\;\; & s_{0|k} = s(k) = s_{1|k-1}^\star \\
        & s_{i+1|k} = \Phi s_{i|k} + B c_{i|k},~i \in \mathbb{I}_{[0,N-1]}\\
        & \tilde{F} s_{i|k} + G c_{i|k} \leq \mathbf{1} - \gamma_{i+k},~i \in \mathbb{I}_{[0,N-1]} \label{eq:indirect_feedback_SMPC_problem_constraint} \\
        & s_{N|k} \in \mathcal{X}_{\mathrm{f}} ^{\mathrm{IF}}\label{eq:eq:indirect_feedback_SMPC_problem_terminal_set}.
    \end{align}
    \label{eq:indirect_feedback_SMPC_problem}
\end{subequations}
Closed-loop operation is similar to RS-MPC: Problem \eqref{eq:indirect_feedback_SMPC_problem} is solved at each $k \in \mathbb{I}_{\geq 0}$, and we apply $u(k) = K x + c_{0|k}^\star$ to the system.
The nominal state $s(k)$ follows the nominal dynamics~\eqref{eq:nominal_state_dynamics} with $s(0) = x(0)$.
We enforce constraints for $i \geq N$ with $c_{i|k}=0$ using the terminal set 
\begin{equation}
    \mathcal{X}_{\mathrm{f}}^{\mathrm{IF}} = \{ s | \tilde{F} \Phi^i s \leq \mathbf{1} - \gammabar, i \in \mathbb{I}_{\geq 0}\},
\end{equation}
where $\gammabar\geq \gamma_k$,~$k\in\mathbb{I}_{\geq 0}$ (App.~\ref{app:tightening_infinite_horizon}). 
As the measured state $x(k)$ enters the optimization problem \eqref{eq:indirect_feedback_SMPC_problem} only through the objective, the scheme is named \textit{indirect feedback SMPC}.
\begin{theorem}
    (adapted from \cite[Thm.~1--2, Cor~1]{Hewing2020indirect})
The SMPC scheme based on Problem~\eqref{eq:indirect_feedback_SMPC_problem} ensures the desired closed-loop properties in Definition~\ref{def:SMPC_good}. 
\end{theorem}
The fact that the nominal state is not reset provides recursive feasibility, as the nominal system develops independently from the error. 
Closed-loop constraint satisfaction follows directly from enforcing constraints \eqref{eq:general_stochastic_constraint} for all $i \in \mathbb{I}_{\geq0}$  with~\eqref{eq:indirect_feedback_SMPC_problem_constraint}.
\subsection{Qualitative analysis}
\label{sec:discussion_qual}
In the following, we compare RS-MPC and IF-SMPC using edge cases that reveal particular shortcomings of either scheme.

\subsubsection{Shortcomings of IF-SMPC}
\label{sec:discussion_qual_IF_worse} 
In the following, we show that IF-SMPC is more conservative than RS-MPC if $p \to 1$ by relating to established robust MPC formulations.

First, note that for $p_j=1$, $j \in \mathbb{I}_{[1,r_{\mathrm{x}}]}$, 
 the constraint tightening constants are equivalent, i.e., $\beta_i=\gamma_i$, compare \eqref{eq:exact_stochastic_tightening_term}, \eqref{eq:robust_tightening_term}, \eqref{eq:robust_stochastic_tightening_term}. 
Furthermore, RS-MPC reduces to a standard robust constraint-tightening MPC~\cite[Sec.~3.2--3.3]{Kouvaritakis2016textbook},\cite{chisci2001robustMPC}.\footnote{\cite[Sec.~3.2--3.3]{Kouvaritakis2016textbook},\cite{chisci2001robustMPC} additionally enforce~\eqref{eq:Robust_Stochastic_MPC_problem_constraints} for $i=0$, which, however, is redundant if $x(0)$ fulfills the state constraints.}
On the other hand, as $k\rightarrow\infty$, the tightening in IF-SMPC~\eqref{eq:indirect_feedback_SMPC_problem_constraint} increases to  $\gammabar\geq \gamma_{i+k}$, which corresponds to the size of the (minimal) robust positive invariant set for the error. 
Hence, the IF-SMPC corresponds to the simple/conservative robust tube MPC scheme in~\cite{Mayne2001tube-mpc_fixed_s0}. 
In particular, compared to a standard robust tube MPC (Mayne et al. (2005)~\cite{Mayne2005regular_tube_MPC}), the fixed nominal initial state $s_{0|k} = s_{1|k-1}^\star$~\eqref{eq:indirect_feedback_SMPC_problem_initialState} reduces the degrees of freedom. 
It is well known in the robust MPC literature, that this simpler approach is more conservative than taking into account the new measured state $x(k)$, in particular whenever the realized disturbance is not the worst-case disturbance~\cite[Sec.~5]{mayne2016robust}, \cite[Sec.~3]{Zanon2021tube_vs_robust}.  
While the independence of the constraints from the measured state and realized disturbances is a key characteristic of IF-SMPC, this does not allow for a ``resetting" of the constraints and thereby makes the approach more conservative as $p\to 1$.

\subsubsection{Shortcomings of RS-MPC}
\label{sec:discussion_qual_rs_worse}
In the following, we discuss cases where the constraint tightening in RS-MPC is significantly more conservative and where this results in a performance deterioration.

By definition, $\gamma_i\leq \beta_i$, since $\gamma_i$ is an exact reformulation (Prop.~\ref{prop:exact_stochastic_constraint}) and $\beta_i$~\eqref{eq:robust_stochastic_tightening_term}  treats only the first disturbance stochastically and the rest robustly (see also proof~\cite[Thm.~8.1]{Kouvaritakis2016textbook}).
There are several cases where this difference becomes extremely large: 
(i) Whenever the disturbance only affects the constraints indirectly through the dynamics ($\tilde{F} D=0$), we have $\gamma_1=0$, and thus, RS-MPC enforces (conservative) worst-case constraints, independent of the probability level $p\in(0,1)$ or the disturbance distribution (see~\cite[Sec.~III.A]{Hewing2020direct_vs_indirect}).
(ii) If $w$ is drawn from a truncated Gaussian with fixed variance, then the conservatism becomes arbitrary large if the support $\mathcal{W}$ is large. 
(iii) Similarly, the difference becomes significant if $p\ll 1$, or more generally, whenever there is a big difference between chance constraints and a purely robust formulation.

Given the discussion in Section~\ref{sec:discussion_qual_IF_worse}, it is not immediately obvious if 
$\gamma_i\ll\beta_i$ implies that RS-MPC is more conservative in closed-loop operation.
Hence, we next provide a clear case where $\gamma_i<\beta_i$ deteriorates the performance of RS-MPC. 
Suppose we wish to stabilize a steady-state which is close to a chance constraint. 
By choosing an LQR feedback $K=K_{\mathrm{LQR}}$, IF-SMPC recovers the optimal LQR performance, assuming the LQR satisfies the chance constraints ($\gamma_i\leq \gammabar\leq 1$)~\cite[Lemma~1]{Hewing2020direct_vs_indirect}. 
RS-MPC can provide the same guarantees if $\beta_i\leq \betabar\leq 1$. 
However, since $\gamma_i<\beta_i$, this condition is in general not satisfied. 
Then, the standard RS-MPC design is infeasible and instead, a suboptimal design choice is required, e.g., stabilizing a steady-state further away from the constraints or trying to decrease $\beta_i$ with a suboptimal tube feedback $K$, resulting in performance deterioration compared to IF-SMPC.

\begin{remark}(Generalizations)
The presented analysis focused on two specific SMPC schemes: RS-MPC and IF-SMPC.
Nonetheless, the qualitative analysis similarly applies more generally to the two SMPC frameworks: \cite{cannon2010stochastic,KouvaritakisCannon2010rs-mpc,cannon2012stochastic,korda2011strongly,Lorenzen2017tightening,kerz2023dataSMPC,Kouvaritakis2016textbook} vs.  ~\cite{farina2013probabilistic,farina2015approach,farina2016model,li2021distributionally,pan2022towards,Hewing2018recovery_mechanism,Koehler2022interpolating,schluter2022stochastic,gruner2022recursively,li2022corrigendum,mark2021stochastic,Hewing2020indirect,mayne2016robust,Hewing2020direct_vs_indirect,mark2021data,hewing2019scenario,muntwiler2022lqg,wang2021recursive}.\footnote{%
The candidate-based re-conditioning in~\cite[Prop.~1]{wang2021recursive} does not seem to suffer from the limitations of IF-SMPC as $p\to 1$.} 
For example, the conservatism of RS-MPC can be reduced by directly enforcing robust recursive feasibility using a robust control invariant set. 
However, similar to the discussion in Section~\ref{sec:discussion_qual_rs_worse}, if the worst-case robust bounds are too conservative,  the design might be infeasible, especially if the desired steady-state is close to a chance constraint.  
Considering IF-SMPC, the conservatism of the fixed initial state constraint $s_{0|k}=s_{1|k-1}^\star$ can be reduced by using a less restrictive interpolating initial state constraint between $s^\star_{1|k-1}$ and the measured state $x(k)$~\cite{Koehler2022interpolating,schluter2022stochastic,gruner2022recursively} (cf. also~\cite{Hewing2018recovery_mechanism,mark2021stochastic,li2022corrigendum} for previous binary initialization strategies). 
While this provides some reduction in conservatism, such an interpolating initial state constraint is still significantly more restrictive than the standard initial state constraint in a robust tube MPC scheme~\cite{Mayne2005regular_tube_MPC}. Hence, the schemes~\cite{Koehler2022interpolating,schluter2022stochastic,gruner2022recursively} are also more conservative than RS-MPC as $p\to 1$.  
\end{remark}

\begin{remark}
\label{rk:input_constraints}
(Input constraints)
In IF-SMPC, no distinction is made between state and input constraints. 
Thus, also probabilistic input constraints can be considered.  
However, for the considered hard input constraints~\eqref{eq:constraint_input}, this treatment can be quite conservative (see discussion Sec.~\ref{sec:discussion_qual_IF_worse} with $p_j=1$). 
By imposing the input constraints for $i=0$ directly based on the measured state $x(k)$ (see~\eqref{eq:Robust_Stochastic_MPC_problem_u0_constraint}), this conservatism can be reduced. 
More generally, IF-SMPC could be modified to implement any robust constraints ($p_j=1$) conditioned on $x(k)$ using the formulas from RS-MPC~\eqref{eq:Robust_Stochastic_MPC_problem} with a different nominal trajectory, while the IF-SMPC formulas are only used for chance constraints ($p_j<1$), which should reduce conservatism. 
\end{remark}


\section{Unifying framework - multi-step SMPC}
\label{sec:multiStep}
Motivated by the limitations exposed in Section~\ref{sec:discussion_qual}, we provide a unifying SMPC framework, which contains these two SMPC formulations as extreme cases.

\subsection{Conceptual idea} 
On a high-level, RS-MPC (Sec.~\ref{sec:RS},\cite{Kouvaritakis2016textbook}) robustly enforces the chance constraints~\eqref{eq:prob_constraint} conditioned on the state one time step in the past. 
Then, recursive feasibility is ensured by accounting for the worst-case disturbance for the rest of the horizon with $\beta_i$~\eqref{eq:robust_stochastic_tightening_term}.  
In contrast, IF-SMPC (Sec.~\ref{sec:IF}, \cite{Hewing2020indirect}) enforces the chance constraints~\eqref{eq:prob_constraint} conditioned on the initial state $x(0)$.  
As a result, constraints are enforced on a nominal state $s$ (independent of the measured state), and a pure stochastic constraint tightening $\gamma_i$~\eqref{eq:exact_stochastic_tightening_term} is used. 

As a natural unification, the proposed framework conditions the chance constraints~\eqref{eq:prob_constraint} on a state (up to) $M\in\mathbb{I}_{\geq 1}$ steps in the past. 
Correspondingly, the first $M$ steps are treated stochastically and the rest robustly, i.e.,  
\begin{align}
\label{eq:beta_tilde} 
&\tilde{\beta}_i  =\gamma_i, ~i\in\mathbb{I}_{[1,M]},\quad  
&\tilde{\beta}_{i+1}=\tilde{\beta}_i+a_{i},~i\in\mathbb{I}_{\geq M}.  
\end{align}
Furthermore, we use a nominal state, which is reset every $M$ steps.
This naturally provides a unified framework, corresponding to IF-SMPC and RS-MPC in case $M\in\{\infty,1\}$, respectively, see discussion in Section~\ref{sec:multistep_discussion}.
   
\subsection{Proposed SMPC algorithm}
Since the nominal state is reset every $M$ steps, we define two nominal states   
\begin{subequations}
\label{eq:MS_initial_state}
\begin{align} 
\label{eq:MS_initial_state_x}
&s(k+1)=
\begin{cases}
x(k+1) & \mathrm{mod}(k+1,M)=0\\
\Phi s(k)+Bc(k)&\text{else} 
\end{cases}\\
&z(k+1)=
\begin{cases}
\Phi s(k)+Bc(k) & \mathrm{mod}(k+1,M)=0\\
\Phi z(k)+Bc(k)&\text{else} 
\end{cases} 
\label{eq:MS_initial_state_s} 
\end{align}
\end{subequations}
which are initialized with $s(0)=z(0)=x(0)$. 
In case\\
$\bmod(k+1,M)\neq 0$, these states follow the nominal dynamics. 
For $\mathrm{mod}(k+1,M)=0$, $s$ is reset to the measured state and $z$ is the nominal prediction based on the measured state $M$  steps in the past (assuming no disturbances $w$). 
In general, $s(k)$ is defined based on $x(k-\mathrm{mod}(k,M))$ and $z(k)$ is defined based on $x(k-M_k)$ with 
\begin{align}
\label{eq:M_k}
M_k:=
\begin{cases}    
M+\bmod(k,M)&k\geq M\\
k& k< M
\end{cases}.
\end{align} 
The corresponding SMPC formulation is given by 
\begin{subequations}
\label{eq:MS}
\begin{align}
\min_{c_{\cdot|k}}  & \;\;\;\mathcal{J}(x(k),c_{\cdot|k})\\
\text{s.t. }&s_{0|k}=s(k)\\
\label{eq:MS_x_dynamics}
& s_{i+1|k} = \Phi s_{i|k} + B c_{i|k} ,~i\in\mathbb{I}_{[0,N-1]}\\
\label{eq:MS_x_constraint}
& \tilde{F} s_{i|k} + G c_{i|k} \leq \mathbf{1} - \tilde{\beta}_{i+\mathrm{mod}(k,M)},~i
\in\mathbb{I}_{[2M-M_k,N-1]}\\ 
\label{eq:MS_x_terminal}& s_{N|k} \in \mathcal{X}^{\mathrm{MS}}_{\mathrm{f}}(\mathrm{mod}(k,M)) \\
\label{eq:MS_s_dynamics}
&z_{0|k}=z(k)\\ 
&z_{i+1|k}=\Phi z_{i|k} + B c_{i|k},~
i\in\mathbb{I}_{[0,2M-1-M_k-1]} \\ 
\label{eq:MS_s_constraint}
& \tilde{F} z_{i|k} + G c_{i|k} \leq \mathbf{1} - \tilde{\beta}_{i+M_k},~i\in\mathbb{I}_{[0,\min\{2M-M_k,N\}-1]}\\
&z_{N|k}\in\mathcal{Z}^{\mathrm{MS}}_{\mathrm{f}}.\label{eq:MS_s_terminal} 
\end{align}
\end{subequations} 
The constraint~\eqref{eq:MS_x_constraint} enforces the chance constraints at time $i+k$ conditioned on $s(k)$ and hence on $x(k-\mathrm{mod}(k,M))$, with $\tilde{\beta}_{i+\mathrm{mod}(k,M)}$. 
The constraint~\eqref{eq:MS_s_constraint} enforces constraints on $i+k$, conditioned on $z(k)$ and hence on $x(k-M_k)$, with the constant $\tilde{\beta}_{i+M_k}$. 
Assuming $M_k=M<N$, the constraints on $z$~\eqref{eq:MS_s_constraint} are imposed on the first $M$ prediction steps and the constraints on $s$~\eqref{eq:MS_x_constraint} on the remaining horizon.
The corresponding terminal sets capture the constraints for $i\geq N$ with $c_{i|k}=0$:\footnote{%
If $N\geq 2M\geq 2M-M_k$, the terminal set $\mathcal{Z}^{\mathrm{MS}}_{\mathrm{f}}$ in~\eqref{eq:MS_s_terminal} can be omitted. 
}
\begin{subequations}
 \label{eq:terminalset_MS}
\begin{align}
 \mathcal{X}^{\mathrm{MS}}_{\mathrm{f}}(k)=&\{s|~\tilde{F}\Phi^i s\leq \mathbf{1}-\tilde{\beta}_{N+i+k},~i\in\mathbb{I}_{\geq 0}\},\\
 \mathcal{Z}^{\mathrm{MS}}_{\mathrm{f}}=&\{z|~\tilde{F}\Phi^i z\leq \mathbf{1}-\tilde{\beta}_{N+i+2M-1},~i\in\mathbb{I}_{\geq 0}\},
\end{align}
\end{subequations}
with $k\in\mathbb{I}_{[0,M-1]}$.  
Since the proposed SMPC framework re-conditions the chance constraints every $M$ steps, it is similar to a multi-step implementation where the first $M$ inputs are directly applied and the optimization problem is only solved every $M$ steps. 
For this reason, we call this approach \textit{multi-step} SMPC (MS-SMPC).  
\begin{theorem}
\label{thm:MS}
The SMPC scheme based on Problem~\eqref{eq:MS} ensures the desired closed-loop properties in Definition~\ref{def:SMPC_good}.
 \end{theorem}
The proof mainly  combines the tools and ideas of RS-MPC and IF-SMPC, compare Appendix~\ref{app:proof} for details.

\subsection{Discussion}
\label{sec:multistep_discussion}
In the following, we show that the proposed formulation indeed recovers RS-MPC (Sec.~\ref{sec:RS}) and IF-SMPC (Sec.~\ref{sec:IF}) 
 for $M=1$ and $M=\infty$, respectively.

For $M=\infty$, the initialization~\eqref{eq:MS_initial_state}, \eqref{eq:M_k} ensures $M_k=k$, and the nominal states $s(k)=z(k)$ are equivalent to the nominal state in IF-SMPC.
 Furthermore, the constraint tightening~\eqref{eq:beta_tilde} is equivalent to IF-SMPC with $\tilde{\beta}_i=\gamma_i$.
 Finally, looking at Problem~\eqref{eq:MS}, note that $\tilde{\beta}_{i+M_k}=\tilde{\beta}_{i+\mathrm{mod}(k,M)}=\gamma_{i+k}$, $s_{\cdot|k}=z_{\cdot|k}$, $\mathcal{Z}^{\mathrm{MS}}_{\mathrm{f}}\subseteq\mathcal{X}^{\mathrm{MS}}_{\mathrm{f}}(k)$. 
Hence, the tightened constraints and terminal set constraint in Problem~\eqref{eq:MS} are equivalent to Problem~\eqref{eq:indirect_feedback_SMPC_problem}.  
 
Next, we consider $M=1$, which yields $\mathrm{mod}(k,M)=0$ and $M_k=1$ (for $k>0$). 
The nominal state $s(k)=x(k)$ is equivalent to RS-MPC, while $z(k)=s_{1|k-1}^\star$.
Furthermore, we trivially have that $\tilde{\beta}_i=\beta_i$ (see~\eqref{eq:beta_tilde},\eqref{eq:robust_stochastic_tightening_term}).
The constraints on $s_{i|k}$ in Problem~\eqref{eq:MS} are posed for $i\geq 2M-M_k=1$ and are hence 
equivalent to Problem~\eqref{eq:Robust_Stochastic_MPC_problem}.\footnote{%
For $k=0$, the constraints~\eqref{eq:MS_x_constraint} on $s$ at $i=2$ are equivalent to the corresponding constraint~\eqref{eq:MS_s_constraint} on $z$.}
The only difference is the constraints on $z$, which are posed for $i\leq \min\{2M-M_k,N\}-1=0$.\footnote{%
Neglecting the terminal set constraint $z_{N|k}\in\mathcal{Z}^{\mathrm{MS}}_{\mathrm{f}}$, which is not needed with $N\geq 2M-M_k=1$.  
} 
The corresponding state constraints ($G_{(j)}=0$) are redundant and can be removed without changing the solution. 
However, there is a difference in the handling of the input constraints ($G_{(j)}\neq 0$) for $i=0$.
In Problem~\eqref{eq:Robust_Stochastic_MPC_problem}, the exact input constraints are  enforced separately utilizing the (known) current state $x(k)$ in~\eqref{eq:Robust_Stochastic_MPC_problem_u0_constraint}. 
Problem~\eqref{eq:MS} robustly enforces the input constraint~\eqref{eq:constraint_input} at time $k$ conditioned on $x(k-1)$, which is more conservative. 
As discussed in Remark~\ref{rk:input_constraints}, this issue can be addressed by explicitly separating the hard input constraints from the chance constraints. 

Overall, the proposed MS-SMPC encompasses both RS-MPC and IF-SMPC and hence allows to flexibly trade off limitations exposed in Section~\ref{sec:discussion_qual} with an appropriate choice of $M\in\mathbb{I}_{\geq 1}$. 
\section{Numerical comparison}
\label{sec:numerical_comparison} 
The following numerical comparison demonstrates the complementary advantages and shortcomings of  RS-MPC and IF-SMPC discussed in Section \ref{sec:discussion_qual}.  
In addition, we show that the proposed MS-SMPC  (Sec.~\ref{sec:multiStep}) can avoid both pitfalls with an appropriate choice of $M\in\mathbb{I}_{\geq 1}$. 
First, we introduce the simulation setup (Sec.~\ref{sec:sim_setup}).
Then, we study two edge cases which demonstrate the limitations of RS-MPC (Sec.~\ref{sec:sim_LQR}) and  IF-SMPC (Sec.~\ref{sec:sim_p_is_1}), respectively.

The MPC problems were formulated with YALMIP~\cite{lofberg2004yalmip} and solved with quadprog (Sec.~\ref{sec:sim_LQR}) and MOSEK~\cite{mosek} (Sec.~\ref{sec:sim_p_is_1})  in Matlab. 
The  code is available online.\footnote{%
gitlab.ethz.ch/ics/MS-SMPC}
\subsection{Simulation setup}
\label{sec:sim_setup}
We consider a $4$th-order integrator system from \cite{Hewing2020direct_vs_indirect} with  
\begin{gather*}
    A =
    \begin{bmatrix}
        1 & T_{\mathrm{s}} & T_{\mathrm{s}}^2/2 & T_{\mathrm{s}}^3/6 \\
        0 & 1   & T_{\mathrm{s}}     & T_{\mathrm{s}}^2/2 \\
        0 & 0   & 1       & T_{\mathrm{s}}     \\
        0 & 0   & 0       & 1
    \end{bmatrix}
    ,~
    B = 
    \begin{bmatrix}
        T_{\mathrm{s}}^4/24 \\
        T_{\mathrm{s}}^3/6  \\
        T_{\mathrm{s}}^2/2  \\
        T_{\mathrm{s}}
    \end{bmatrix}
    ,~
    D = B
\end{gather*}
and $T_{\mathrm{s}} = 0.1$.
The disturbances $w(k)$ are uniformly distributed over the set $\mathcal{W}=[-4,4]$ and the input constraint is $u(k)\in[-20,20]$.  
This system is interesting as the effect of the disturbance $w$ and the control input $u$ on the first state $x_{(1)}$ is only apparent over longer horizons (see also~\cite{Hewing2020direct_vs_indirect}).
Hence, we consider a chance constraint on the first state: $\prob{x_{(1)}(k) \leq 0.1} \geq p$, with some probability level $p$ specified later.  
We use condition~\eqref{eq:Robust_Stochastic_MPC_problem_u0_constraint} in IF-SMPC and MS-SMPC to implement the current input constraints ($i=0$) in a non-conservative fashion, similar to RS-MPC (Rk.~\ref{rk:input_constraints}). 
All simulations are performed for $10^3$ realizations over 300 steps, with initial state $x(0) = [0~0~0~0]^\top$. 
The average cost $\ell_{\mathrm{avg}}$~\eqref{eq:performance} and average satisfaction of the chance constraints are approximated using only the interval $k \in \mathbb{I}_{[50,299]}$.

\subsection{Case 1 - performance limitations of RS-SMPC}
\label{sec:sim_LQR} 
First, we present an example where IF-SMPC has better performance than RS-MPC. We choose probability level $p = 0.7$, and  a stage cost $\ell(x,u) = \| x \|_Q^2 + \| u \|_R^2$ 
where $Q = \text{diag} \left( Q_{11}, 0, 0, 0 \right)$ and $R = 0.1$, with $Q_{11} = 1.32$.   
Further, we select prediction horizon $N=75$ and multi-step horizon $M=35$.

\begin{figure}[t]
    \centering 
    \includegraphics[width=0.4\textwidth]{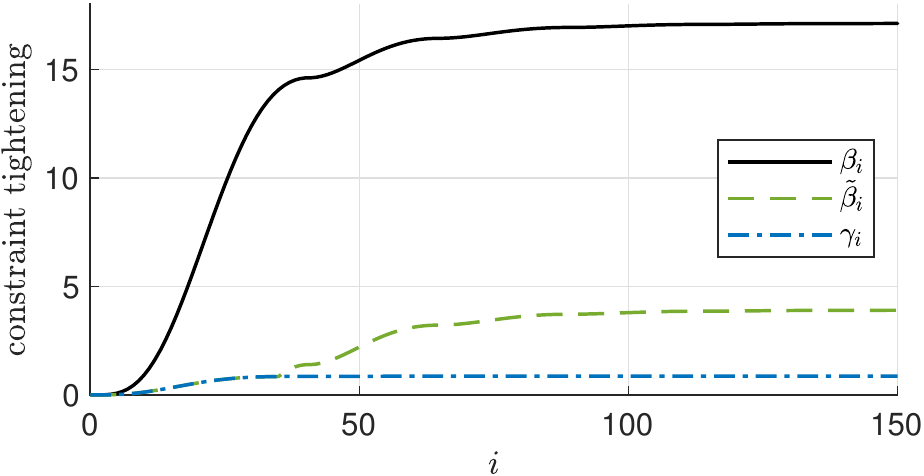}

    \caption{Comparison of state constraint tightening with tube controller $K_{\mathrm{LQR}}$.}
    \label{fig:LQR_constraint_tightening}
\end{figure}

We consider an optimal LQR state feedback gain $K_{\mathrm{LQR}}$ (tube controller). 
Figure~\ref{fig:LQR_constraint_tightening} shows the resulting constraint tightening $\beta_i$ (RS-MPC), $\tilde{\beta}_i$ (MS-SMPC), and $\gamma_i$ (IF-SMPC) with $K_{\mathrm{LQR}}$, 
indicating large conservatism of the robust-stochastic tightening $\beta_i$.  
Furthermore, we have $\gammabar\approx 0.9$, while $\tildebetabar \geq \lim_{i \to \infty} \tilde{\beta}_i\approx 4$, and $\betabar \geq \lim_{i \to \infty} \beta_i>15$.  
Therefore, the tightened constraints are empty, rendering the MS-SMPC and RS-MPC designs infeasible. 
To allow for application of MS-SMPC and RS-MPC, we choose more aggressive, suboptimal tube controllers $K_{\mathrm{MS}}$ and $K_{\mathrm{RS}}$ with $Q_{11}$ increased to $Q_{\mathrm{MS},11}=16.6$ and $Q_{\mathrm{RS},11}=740$, respectively, which results in sufficiently small tightening $\betabar \approx 0.9$ and $ \tildebetabar \approx 0.9$.

\subsubsection*{Results} 
\begin{figure}[t]
    \centering 
     \includegraphics[width=0.45\textwidth]{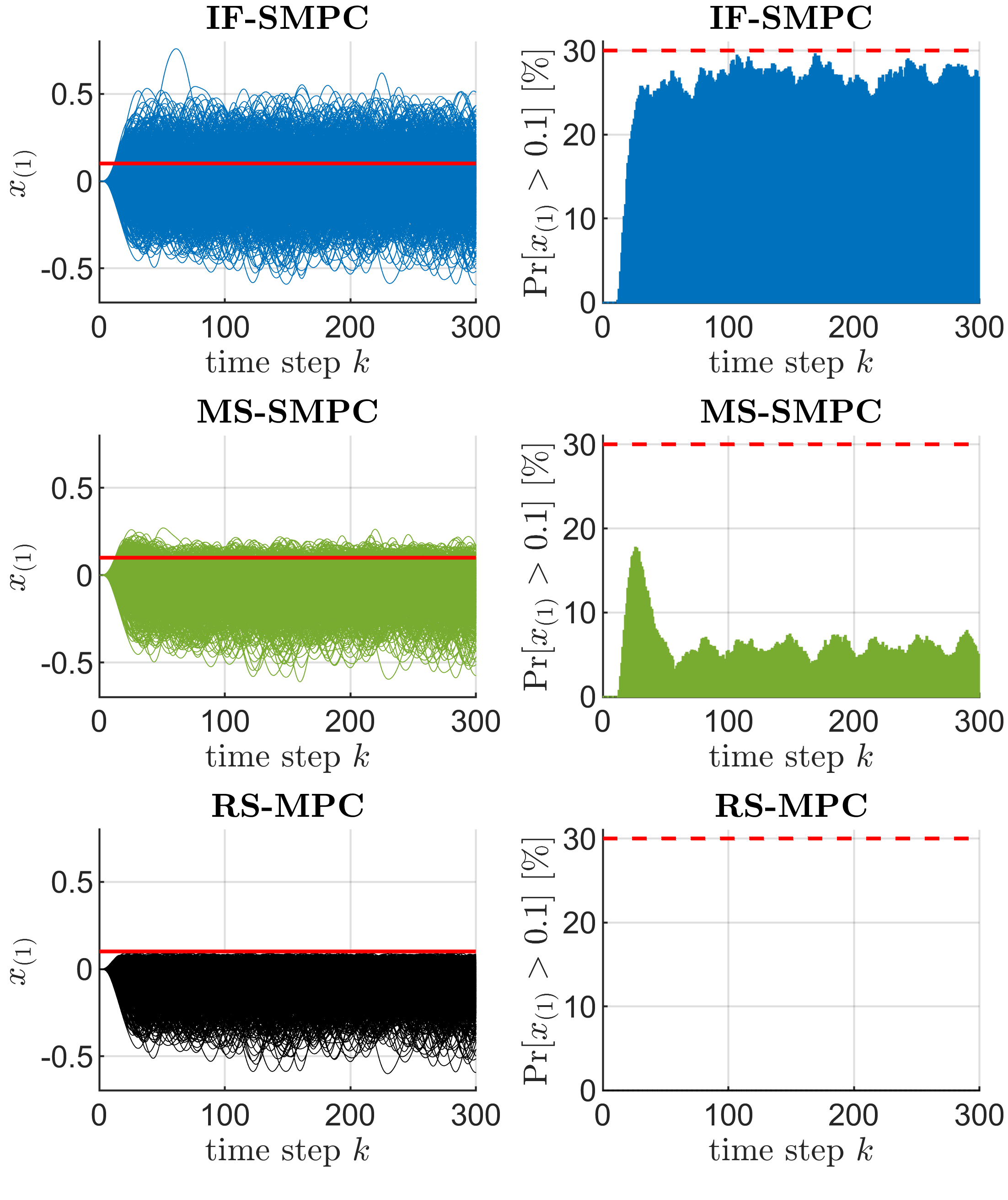}
\caption{Case 1: State trajectories (left) and probability of constraint violation (right) of IF-SMPC (blue), MS-SMPC (green, $M = 35$), and RS-MPC (black).
    The solid red lines depict the state constraints; the dashed red lines indicate the violation level $1-p$.}
    \label{fig:LQR_state_trajectories_violations}
\end{figure}

\begin{figure}[t]
    \centering 
    \includegraphics[width=0.4\textwidth]{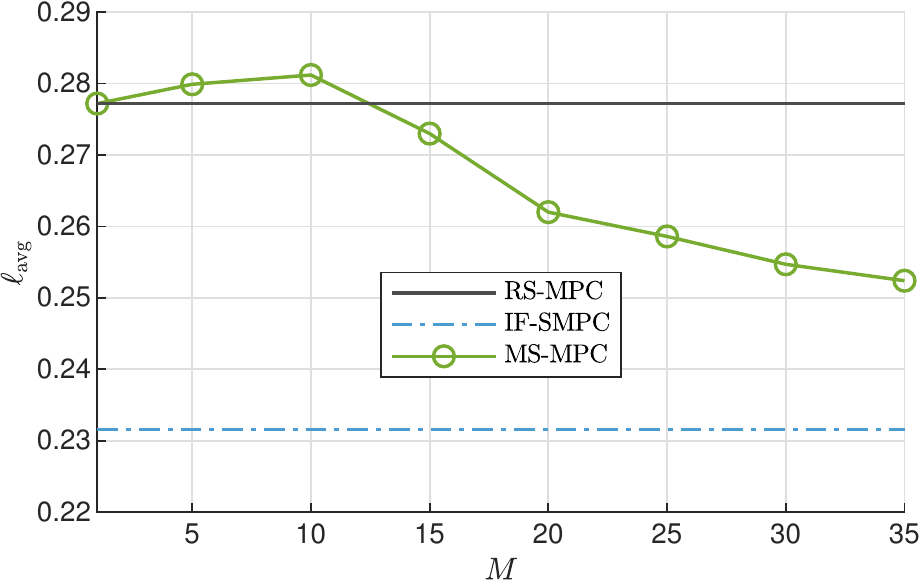}
       \caption{Case 1: Performance of MS-SMPC for different multi-step horizons $M$ in comparison to RS-MPC and IF-SMPC.}\vspace{-4mm}
    \label{fig:LQR_MS_performance_over_M}
\end{figure}

The state trajectories and corresponding constraint violations are presented in 
Figure~\ref{fig:LQR_state_trajectories_violations}.  
Table~\ref{tab:case_1}  details performance (normalized w.r.t. LQR performance) and constraint violation probability of all SMPC schemes and their respective tube controllers. 
Figure~\ref{fig:LQR_MS_performance_over_M} shows the performance of MS-SMPC for different multi-step horizons $M$.\footnote{%
For each $M$, we compute the least conservative feedback $K$ ensuring $\tildebetabar\approx 0.9$.} 

\begin{table}[ht]
\centering
    \begin{tabular}{ l  c  c}
        \hline
         & ${\ell_{\mathrm{avg}}}$ & {Violation probability} \\
        \hline  
        ${K_{\mathrm{LQR}}}$ & $1.00$ & $27.08 \%$ \\   
        ${K_{\mathrm{MS}}}$ & $1.26$ & $4.07 \%$ \\  
        ${K_{\mathrm{RS}}}$ & $2.29$ & $0.00 \%$ \\   
        \hline
        {IF-SMPC with ${K_{\mathrm{LQR}}}$} & $1.00$ & $27.08 \%$ \\  
        {MS-SMPC with ${K_{\mathrm{MS}}}$} & $1.09$ & $5.75 \%$ \\  
        {RS-MPC with ${K_{\mathrm{RS}}}$} & $1.20$ & $0.00 \%$ \\  
        \hline
    \end{tabular} 
\caption{Case 1: Comparison - performance and violation probability.}
\label{tab:case_1}
\end{table}


We observe a significant performance improvement of IF-SMPC over RS-MPC.  
Due to  $\tilde{F}_{(j)} D  \approx 0$ for the state constraint $j=1$, the constraint on $x_{(1)}$ is treated virtually robustly in the RS-MPC case (Sec.~\ref{sec:discussion_qual_rs_worse}), resulting in practically no constraint violations and higher cost, while the probability of constraint violation for IF-SMPC is close to the specified $1-p=30\%$.
Further, IF-SMPC matches performance and constraint violations of the LQR. 
For $M = 35$, MS-SMPC achieves performance and constraint violations in between RS-MPC and IF-SMPC.
As expected, Figure~\ref{fig:LQR_MS_performance_over_M} indicates that for $M=1$, MS-SMPC matches the performance of RS-MPC while for $M \to \infty$, we observe convergence to the performance of IF-SMPC (Sec.~\ref{sec:multistep_discussion}).

\subsection{Case 2 - performance limitations of IF-SMPC}
\label{sec:sim_p_is_1}
In the following, we present an  example where RS-MPC has better performance than IF-SMPC.
To this end, we minimize an economic cost objective with stage cost $\ell(x) = -x_{(1)}+0.1$.  
In consequence, $x_{(1)}$ is maximized and the state constraint is always active. 
We choose probability level $p = 0.9$, tube control gain $K = [-55.45,~-51.22,~-23.65,~-6.54]$, prediction horizon $N=30$, and multi-step horizon $M=2$.   

\subsubsection*{Results}
\begin{figure}[t]
\centering\includegraphics[width=0.45\textwidth]{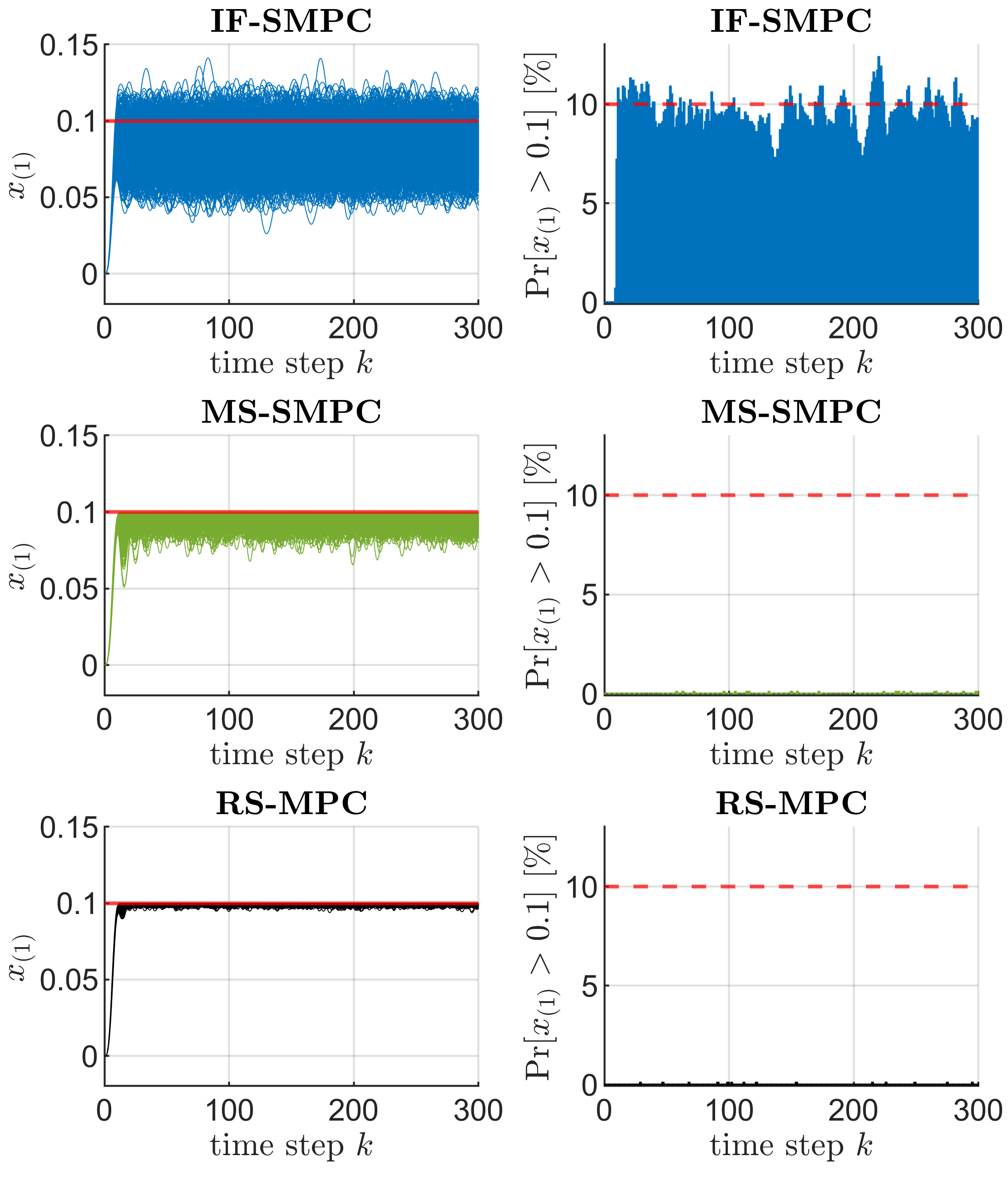}
    \caption{Case 2: State trajectories (left) and probability of constraint violation (right) of IF-SMPC (blue), MS-SMPC (green, $M = 2$), and RS-MPC (black).
    The solid red lines depict the state constraints; the dashed red lines indicate the violation level $1-p$.}
    \label{fig:p1_state_trajectories_violations}
\end{figure}

State trajectories and constraint violations are shown in  
Figure~\ref{fig:p1_state_trajectories_violations}. 
Performance and  constraint violation probability are compared in Table~\ref{tab:case_2}. 
Similarly to Case~1, RS-MPC results in virtually no constraint violations due to the (approximately) robust disturbance treatment by RS-MPC in this example (Secs. \ref{sec:discussion_qual_rs_worse}, \ref{sec:sim_LQR}).
Nonetheless, RS-MPC has significantly better performance than IF-SMPC, as the resetting of the nominal state in RS-MPC reduces conservatism considerably. 
Again, MS-SMPC can achieve performance in between RS-MPC and IF-SMPC.  
\begin{table}[ht]
\centering
    \begin{tabular}{ l  c  c}
        \hline
         & ${\ell_{\mathrm{avg}} \cdot 10^{2}}$ & {Violation probability} \\
        \hline
        {IF-SMPC} & $1.71$ & $9.485 \%$ \\ 
        {MS-SMPC} & $0.41$ & $0.008 \%$ \\ 
        {RS-MPC} & $0.06$ & $0.005 \%$ \\  
        \hline
    \end{tabular}
\caption{Case 2: Comparison - performance and violation probability.}
\label{tab:case_2}
\end{table}

In summary, we have seen that there can be significant performance differences between RS-MPC and IF-SMPC and either scheme  can be superior, depending on the specific problem.   
Furthermore, we have seen that MS-SMPC unifies RS-MPC and IF-SMPC and can alleviate the respective limitations with a suitable choice of $M$.
\section{Conclusion}
\label{sec:conclusion}
We investigated SMPC schemes with desired closed-loop guarantees. 
We categorized the corresponding literature in two separate frameworks 
and we provided a qualitative analysis, highlighting some of the intrinsic features and limitations of these two approaches.
We also provided a numerical comparison that support these findings. 
As a separate contribution, we derived a novel SMPC framework that naturally unifies these two approaches. 
The proposed unifying SMPC framework forms a natural basis for future investigations.

\setstretch{0.975}
\bibliographystyle{IEEEtran}  
\bibliography{Bibliography} 
\clearpage
\appendix
%
\subsection{Proof - Theorem~\ref{thm:MS}}
\label{app:proof}
 \begin{proof}
\textbf{Part I. Recursive feasibility:} First, we show feasibility at time $k+1$, given feasibility at time $k\in\mathbb{I}_{\geq 0}$, using the candidate solution $c_{i|k+1}=c_{i+1|k}^\star$, with $c_{i|k}^\star=0$,  and $s^\star_{i|k},z^\star_{i|k}$ according to~\eqref{eq:MS_s_dynamics}
,\eqref{eq:MS_x_dynamics} for $i\geq N$. 
Note that $s^\star_{i|k},z_{i|k}^\star,c_{i|k}^\star$ also satisfy the constraints~\eqref{eq:MS_x_constraint},\eqref{eq:MS_s_constraint} for $i\in\mathbb{I}_{\geq N}$ due to the terminal set~\eqref{eq:MS_s_terminal}, \eqref{eq:MS_x_terminal}, \eqref{eq:terminalset_MS}. 
We show feasibility of the constraints~\eqref{eq:MS_x_constraint}, \eqref{eq:MS_s_constraint}  using a case distinction. 
The two cases are comparable to the proof/method in IF-SMPC and RS-MPC, respectively.\\
\textbf{Case a): }Suppose $\mathrm{mod}(k+1,M)\neq 0$, yielding
$s_{i|k+1}=\overline{s}_{i+1|k}^\star$, $z_{i|k+1}=z_{i+1|k}^\star$,  $i\in\mathbb{I}_{\geq 0}$. 
Furthermore,  $i+\mathrm{mod}(k+1,M)=i+1+\mathrm{mod}(k,M)$ and 
hence, for this candidate at time $k+1$, the constraints \eqref{eq:MS_s_constraint} and \eqref{eq:MS_x_constraint} are equivalent to the constraints satisfied at time $k$, similar to IF-SMPC~\cite{Hewing2020indirect}. The same holds for the terminal set constraints~\eqref{eq:MS_s_terminal},\eqref{eq:MS_x_terminal}. \\
\textbf{Case b): }
Suppose $\mathrm{mod}(k+1,M)=0$, and hence 
$1+\mod(k,M)=M=M_{k+1}$ (since $k+1>0$).
In this case, $z(k+1)=s^\star_{1|k}$ and hence $z_{i|k+1}=s_{i+1|k}^\star$. 
Thus, for $i\in\mathbb{I}_{[0,M-1]}$, conditions~\eqref{eq:MS_s_constraint} hold with
\begin{align*}
& \tilde{F} z_{i|k+1} + G c_{i|k+1} 
= \tilde{F} s^\star_{i+1|k} + G c^\star_{i+1|k} \\
\stackrel{\eqref{eq:MS_x_constraint}}{\leq} &
\mathbf{1} - \tilde{\beta}_{i+1+\text{mod}(k,M)}
=\mathbf{1} - \tilde{\beta}_{i+M_{k+1}}.
\end{align*} 
It holds that 
\begin{align*}
&s_{0|k+1}\stackrel{\eqref{eq:MS_initial_state_x}}{=}x(k+1) 
\stackrel{\eqref{eq:MS_initial_state_x}}{=}s^\star_{1|k}+\sum_{l=0}^{M-1}\Phi^lDw(k-l)
\end{align*}
and hence 
    \begin{align*} 
s_{i|k+1}=&s_{i+1|k}^\star+\sum_{j=0}^{M-1}\Phi^{j+i}w(k-j), ~i\in\mathbb{I}_{\geq 0}.
\end{align*}
The constraints~\eqref{eq:MS_x_constraint} for $i\geq M$ hold with 
\begin{align*}
&\tilde{F}s_{i|k+1}+Gc_{i|k+1}\\ 
=&\tilde{F}s_{i+1|k}^\star+G c_{i+1|k}^\star+\sum_{j=0}^{M-1}\tilde{F}\Phi^{j+i}w(k-j)
\\
\stackrel{\eqref{eq:robust_tightening_term}}{\leq} &\tilde{F}s_{i+1|k}^\star+G c_{i+1|k}^\star+
\sum_{l=i}^{i+M-1} a_{l} \\ 
\stackrel{\eqref{eq:MS_x_constraint}}{\leq}& 1-\tilde{\beta}_{i+1+\mathrm{mod}(k,M)}+\sum_{l=i}^{i+M-1} a_l  
\stackrel{\eqref{eq:beta_tilde}}{=}1-\tilde{\beta}_{i+\mathrm{mod}(k+1,M)},
\end{align*}
where last equality used $i+1+\mathrm{mod}(k,M)=M+i$ and $i+\mathrm{mod}(k+1,M)=i$. 
Similarly, the terminal set constraints~\eqref{eq:MS_x_terminal} and \eqref{eq:MS_s_terminal} hold with
\begin{align*}
&\tilde{F}\Phi^iz_{N|k}= \tilde{F}\Phi^i s^*_{N+1|k}\\
\leq &\mathbf{1}-\tilde{\beta}_{N+i+1+\mathrm{mod}(k,M)}=\mathbf{1}-\tilde{\beta}_{N+i+M}
\leq \mathbf{1}-\tilde{\beta}_{N+i+2M-1}
\end{align*}
and
\begin{align*}
&\tilde{F}\Phi^is_{N|k}\leq \tilde{F}\Phi^i s^*_{N+1|k}+\sum_{l=i}^{i+M-1}a_{N+l}\\
\leq &\mathbf{1}-\tilde{\beta}_{N+i+1+\mathrm{mod}(k,M)}+\sum_{l=i}^{i+M-1}a_{N+l}=\mathbf{1}-\tilde{\beta}_{i+N}. 
\end{align*} 
\textbf{Part II. Chance constraint satisfaction: }
For any $k\in\mathbb{I}_{\geq 0}$, feasibility of Problem~\eqref{eq:MS}, $N\geq 1$, and $M\geq 1$ ensure that condition~\eqref{eq:MS_s_constraint} holds for $i=0$. This condition ensures the individual chance constraint~\eqref{eq:general_stochastic_constraint} conditioned on the state $x(k-M_k)$, see~\eqref{eq:MS_initial_state_s} and Prop.~\ref{prop:exact_stochastic_constraint}. 
Satisfaction of \eqref{eq:general_stochastic_constraint} for the closed-loop system follows from recursive feasibility, similar to RS-MPC~\cite[Thm.~7.1]{Kouvaritakis2016textbook}.
\\ 
\textbf{Part III. Performance bound: }
The candidate solution and cost function in Problem~\eqref{eq:MS} are equivalent to~\cite{Hewing2020indirect,Koehler2022interpolating}.
Hence, it holds that
\begin{align*}
&\expect{\mathcal{J}(x(k+1),c_{\cdot|k+1})}\\
\leq& \mathcal{J}(x(k),c_{\cdot|k}^\star)-\ell(x(k),u(k))+\tr{P_{\mathrm{f}}\Sigma_{\mathrm{w}}}.    
\end{align*}
Using the assumed lower bound on $\mathcal{J}$ and a telescopic sum yields~\eqref{eq:performance}.
\end{proof}
 
\subsection{Constraint tightening computation}
\label{app:tightening_infinite_horizon}
In the following, we discuss the computation of the constraint tightening constants $\gamma_i$  in~\eqref{eq:exact_stochastic_tightening_term}.
First, we recap the scenario approach. 
Then, we discuss the computation of over-approximations for large horizons $k\geq \bar{k}$.

\subsubsection*{Scenario approach}
In the following, we briefly recap how Problem~\eqref{eq:exact_stochastic_constraint} can be (approximately) solved for general probability distributions using sampling, similar to~\cite[Sec.~V.A]{Lorenzen2017tightening} (see also~\cite{campi2011sampling}). 
We specify a desired confidence level $\delta\ll 1$ (e.g., $10^{-6}$), the number of used samples $N_{\mathrm{s}}\gg 1$, 
and compute the number of discarded samples $r_{(j)}\in\mathbb{I}_{\geq 0}$ such that
\begin{align}
r_{(j)}\leq (1-p_{(j)})N_{\mathrm{s}}-\sqrt{2(1-p_{(j)})N_{\mathrm{s}}\ln(1/\delta)}.    
\end{align}
Then, we sample $N_{\mathrm{s}}$ disturbances sequences $w^{(l)}_{[0,k-1]}\in\mathcal{W}^k$, $l\in\mathbb{I}_{[1,N_{\mathrm{s}}]}$ and compute the resulting error $e^{(l)}(k)$ according to~\eqref{eq:error_dynamics}.  
We order the samples $l$ in ascending order for $\tilde{F}_{(j)}e^{(l)}$. 
Finally, we set $\gamma_{(j)}$ based on the $1-r_{(j)}/N_{\mathrm{s}}$ quantile of the samples:
\begin{align}
\label{eq:gamma_sampling}
\gamma_{k,(j)}=\tilde{F}_{(j)}e^{(l)}(k),\quad l=N_{\mathrm{s}}-r_{(j)}.
\end{align}
Note that as $N_{\mathrm{s}}\to\infty$, we get $r_{(j)}/N_{\mathrm{s}}\approx 1-p_{(j)}$ for any $\delta>0$. 
\begin{proposition}(adapted from \cite[Prop.~5]{Lorenzen2017tightening})
With confidence $1-\delta$, the constant $\gamma_{k,(j)}$ according to~\eqref{eq:gamma_sampling}
 satisfies~\eqref{eq:exact_stochastic_tightening_term_2}. 
\end{proposition}

In the numerical results in Section~\ref{sec:numerical_comparison}, we set $N_{\mathrm{s}}=10^6$, $\delta=10^{-4}$. 
For $p=0.9$, this results in $r/N_{\mathrm{s}}=9.86\%<10\%=1-p$. 
We additionally over-approximate the constants a posterior to ensure monotonicity of $\gamma_i$ w.r.t. $i$. 

 \subsubsection*{Over-approximation for infinite-horizon}
In IF-SMPC, the tightened constraints~\eqref{eq:indirect_feedback_SMPC_problem_constraint} depend on $\gamma_{i+k}$. 
Thus, as $k\rightarrow\infty$, this requires the computation of $\lim_{k\rightarrow\infty}\gamma_k$, in which case the previously outlined scenario approach becomes intractable. 

\begin{figure}[t]
    \centering
    \includegraphics[width=0.45\textwidth]{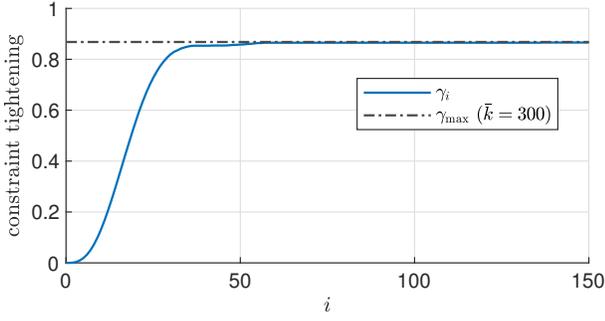}
    \caption{Comparison of state constraint tightening $\gamma_k$ and over-approximation $\gammabar$ with $K_{\mathrm{LQR}}$ from Sec.~\ref{sec:sim_LQR}.}
    \label{fig:LQR_gamma_vs_gamma_bar}
\end{figure}

To mitigate this issue, we derive an upper bound $\gammabar\geq \gamma_i$ for all $i\in\mathbb{I}_{\geq 0}$. 
The following derivation is inspired by the upper bound on $\beta_i$ presented in~\cite[Lemma 8.2]{Kouvaritakis2016textbook}.
First, note that $\gamma_{i+1}-\gamma_{i}\leq a_i$, with $a_i$ according to~\eqref{eq:robust_constraint}, i.e., treating the last disturbance robustly is more conservative than the exact stochastic constraint tightening. 
Hence, by computing $\gamma_i$ for all $i\in\mathbb{I}_{[1,\bar{k}]}$ with some user-chosen $\bar{k}\gg 1$ we have:
$\gamma_i\leq \gamma_{\bar{k}}+\sum_{l=\bar{k}}^\infty a_l$ for all $i\in\mathbb{I}_{>\bar{k}}$. 
Given that $\Phi$ is Schur stable, it holds that $\|\Phi^k \|\leq C\rho^k$, $\forall k\in\mathbb{I}_{\geq 0}$ with some constant $C\geq 1$ and an exponential decay rate $\rho\in[0,1)$, e.g., based on the largest absolute eigenvalue of $\Phi$. 
Hence, for any $j\in\mathbb{I}_{[1,r]}$ it holds that
\begin{align*} 
&\sum_{i=\bar{k}}^\infty a_{i,(j)}
\stackrel{\eqref{eq:robust_constraint}}{=}\sum_{i=\bar{k}}^\infty \max_{w_i\in\mathcal{W}}\tilde{F}_{(j)} \Phi^i D w_i  \\
\leq &\sum_{i=\bar{k}}^\infty \|F_{(j)}\| \|\Phi^i\| \max_{w\in\mathcal{W}}\|Dw\|\\
\leq&  \underbrace{\|F_{(j)}\|\max_{w\in\mathcal{W}}\|Dw\|C}_{=:\tilde{C}_{(j)}}\sum_{i=\bar{k}}^\infty\rho^i
=\tilde{C}_{(j)}\dfrac{\rho^{\bar{k}}}{1-\rho},
\end{align*}
where the last equation used the geometric series. 
In combination, we obtain
\begin{align}
 \label{eq:gamma_bar}   
\gammabar:=\gamma_{\bar{k}}+\tilde{C}\dfrac{\rho^{\bar{k}}}{1-\rho}.
\end{align}
Note that $\gamma_{i+1}\geq \gamma_i$ (monotonicity) implies that $\gammabar\geq \gamma_\infty\geq \gamma_i$, $\forall i\in\mathbb{I}_{\geq 0}$
\begin{proposition}
It holds that $\gammabar\geq \gamma_i$ for any $i\in\mathbb{I}_{\geq 0}$. 
\end{proposition}
Note that the conservatism of setting $\gamma_i=\gammabar$, $i\geq \bar{k}$ converges exponentially to zero by increasing $\bar{k}$ with $\rho^{\bar{k}}$.  
A corresponding illustration of this over-approximation can be seen in Figure~\ref{fig:LQR_gamma_vs_gamma_bar}.  
   
\end{document}